\documentclass[twocolumn]{article}
\usepackage[T1]{fontenc}
\usepackage[utf8]{inputenc}
\usepackage{fullpage}
\usepackage{graphicx,pgf,tikz} % Graphics
\usepackage{environ} % For making a command to resize tikz figures
\usepackage{subfigure}
\usepackage{amsmath,amssymb,amsfonts,amsthm,nicefrac} % Maths stuff
\usepackage{algorithmicx,algorithm,eqparbox,array}
\usepackage[noend]{algpseudocode}
\usepackage{multirow,multicol,booktabs} % Multirowy, Multicolumny and booktabby tables
\usepackage{enumitem} % For better control of enumerate and itemize
\usepackage{lipsum,mwe} % Dummy text and image
\usepackage{hyperref,url}
\usepackage{times}

\algnewcommand{\LineComment}[1]{\State // #1}
\usetikzlibrary{matrix,arrows,decorations.pathmorphing,shapes,calc,backgrounds,fit}
\tikzset{node style ge/.style={rectangle}}

\makeatletter
\newsavebox{\measure@tikzpicture}
\NewEnviron{scaletikzpicturetowidth}[1]{%
  \def\tikz@width{#1}%
  \begin{lrbox}{\measure@tikzpicture}%
  \BODY
  \end{lrbox}%
  \pgfmathparse{#1/\wd\measure@tikzpicture}%
  \BODY
}
\makeatother

\graphicspath{{pictures/}}
\begin{document}
\title{Multitask Learning on Graph Neural Networks: Learning Multiple Graph Centrality Measures with a Unified Network\footnote{The final authenticated publication is availale online at \url{https://doi.org/10.1007/978-3-030-30493-5_63}}}
\author{
    Pedro H.C. Avelar \href{https://orcid.org/0000-0002-0347-7002}{\includegraphics[height=10pt]{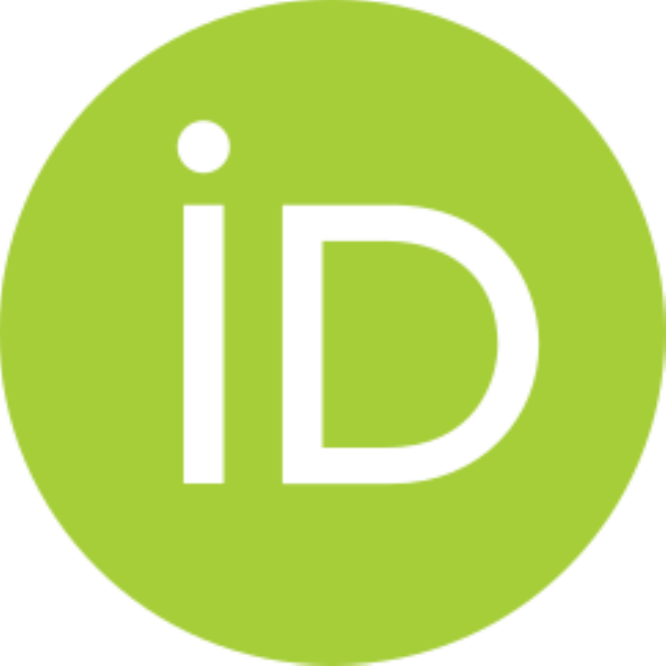}}
    \and Henrique Lemos \href{https://orcid.org/0000-0003-0236-1291}{\includegraphics[height=10pt]{orcid.pdf}}
    \and Marcelo O.R. Prates \href{https://orcid.org/0000-0002-5576-7060}{\includegraphics[height=10pt]{orcid.pdf}}
    \and Luis C. Lamb \href{https://orcid.org/0000-0003-1571-165X}{\includegraphics[height=10pt]{orcid.pdf}} \\\{phcavelar,hlsantos,morprates,lamb\}@inf.ufrgs.br
    \\Institute of Informatics, UFRGS, Porto Alegre, Brazil
}
\date{}

\maketitle

\begin{abstract}
The application of deep learning to symbolic domains remains an active research endeavour. Graph neural networks (GNN), consisting of trained neural modules which can be arranged in different topologies at run time, are sound alternatives to tackle relational problems which lend themselves to graph representations.
In this paper, we show that GNNs are capable of multitask learning, which can be naturally enforced by training the model to refine a single set of multidimensional embeddings $\in \mathbb{R}^d$ and decode them into multiple outputs by connecting MLPs at the end of the pipeline.  
We demonstrate the multitask learning capability of the model in the relevant relational problem of estimating network centrality measures, focusing primarily on producing rankings based on these measures, i.e. is vertex $v_1$ more central than vertex $v_2$ given centrality $c$?.
We then show that a GNN can be trained to develop a \emph{lingua franca} of vertex embeddings from which all relevant information about any of the trained centrality measures can be decoded. 
The proposed model achieves $89\%$ accuracy on a test dataset of random instances with up to 128 vertices and is shown to generalise to larger problem sizes. The model is also shown to obtain reasonable accuracy on a dataset of real world instances with up to 4k vertices, vastly surpassing the sizes of the largest instances with which the model was trained ($n=128$). Finally, we believe that our contributions attest to the potential of GNNs in symbolic domains in general and in relational learning in particular.
\end{abstract}

\section{Introduction}

A promising technique for building neural networks on symbolic domains is to enforce permutation invariance by connecting adjacent elements of the domain of discourse through neural modules with shared weights which are themselves subject to training. By assembling these modules in different configurations one can reproduce each graph's structure, in effect training neural components to compute the appropriate messages to send between elements. The resulting architecture can be seen as a message-passing algorithm where the messages and state updates are computed by trained neural networks. This model and its variants are the basis for several  architectures such as message-passing neural networks \cite{gilmer2017neural}, recurrent relational networks \cite{palm2017recurrent}, graph networks \cite{battaglia2018relational}, graph convolutional networks \cite{kipf2017convolutional} and graph neural networks (GNN) \cite{gori2005gnn,scarselli2005graph,scarselli2009graph} whose terminology we adopt.

GNNs have been successfully employed on both theoretical and practical combinatorial domains, with \cite{palm2017recurrent} showing how they can tackle Sudoku puzzles, \cite{selsam2018learning} developing a GNN which is able to predict the satisfiability of CNF boolean formulas, \cite{gilmer2017neural} using them to predict quantum properties about molecules and \cite{santoro2017simple} using it outside of the context of graphs for relational question answering about pictures. Some of these approaches can even extend their computation to more iterations of message-passing, showing that GNNs can not only learn from examples, but may learn to reason in an iterative fashion.

With this in mind, we turn to the relevant problem of approximating centrality measures on graphs, a combinatorial problem with very relevant applications in our highly connected world, including the detection of power grid vulnerabilities \cite{liu2018power}, influence inside interorganisational and collaboration networks \cite{chen2017research}, social network analysis \cite{kim2018social}, pattern recognition on biological networks \cite{estrada2018bio} among others. This work concerns itself with whether a neural network can approximate centrality measures solely from a network's structure.

The remainder of the paper is structured as follows. First we provide a survey of related work on approximating centrality measures with neural networks. Then, we present the basic concepts of centrality measures used in this paper, introduce some GNN-based models for approximating and learning the relations between centralities in graphs, describe our experimental evaluation, and verify the models' generalisation and interpretability. Finally, we conclude with the contributions and shortcomings of our work, and point out direction for further research.

\section{Related Work}

With regard to our problem at hand, related work can be divided into two fronts. One is that of neural networks approximating centralities on graphs, and the other is the related work on modelling node-level embeddings with graph neural networks. On this first aspect, there are works such as \cite{grando2015estimating}, \cite{grando2016approximating}, \cite{grando2018computing}, \cite{grando2018machine} and \cite{kumar2015neural}, all of which uses neural networks to estimate centrality measures. However, in \cite{grando2015estimating}, \cite{grando2016approximating} and \cite{grando2018machine} they use a priori knowledge of other centralities to approximate a different one, on \cite{grando2018computing} they also produce a ranking of the centrality measures, but do so using the degree and eigenvector centralities as input, and in \cite{grando2016approximating} local features such as number of vertices in a network, number of edges in a network, degree and the sum of the degrees of all of the vertex's neighbours are used. These contributions differ from ours in that we feed our neural network solely with the network structure -- that is, they use a simple MLP which receives numeric information about a specific node and outputs an approximation of one desired centrality measure, while our builds a message-passing procedure with only the network structure and no numeric information whatsoever. \cite{scarselli2005graph} fits both fronts, since it also uses GNNs to compute rankings for the PageRank centrality measure for a single graph, and does not focus on other centralities nor analyses the transfer between centralities.

In \cite{perozzi2014deepwalk} latent representations of networks are learned, akin to our models, but their work does not focus on predicting centrality measures per-se. The works in \cite{duran2017learning,hamilton2017inductive,tang2015line} also concern themselves with creating node embeddings, but again their work does not focus on centrality measures and use features other than the network structure as inputs, specifically \cite{tang2015line} trains its predictor for each specific graph, while also using the node's degree as information. Most of these, however, either train their models using the same network distributions which they use for evaluation, or learn embeddings specific to each graph, while we learn an algorithmic procedure from synthetic data distributions and try to extrapolate to other distributions, which makes direct comparison somewhat difficult.  On the other hand, works such as \cite{li2018learning,you2018graphrnn,you2018graph} work on learning generative models for graph generation, the first and the latter applying their work to the generation of chemical compounds, this can be seen as a step forward in modelling parameters for graphs, but their analyses do not study the relationships between the latent space and the characteristics of the networks itself. For a survey of the area of Graph Neural Networks in general, one can look at \cite{gilmer2017neural,battaglia2018relational,zhang2018deep}.

Given this, our contributions are: (1) The experimental analyses of different graph neural network approaches to ranking nodes and how well they scale in a multitask environment -- in contrast with non-GNN models and those which were not used for multitasking. (2) The proposal of using a learned native comparison method -- instead of approximating the metrics themselves to produce rankings. (3) The training of a model which predicts values solely from the network structure, which is trained in graphs altogether different from those where the model is tested -- instead of optimising the model for a specific graph or training it in graphs from the same distribution as the target ones. (4) We briefly analyse the relationship behind the learned computed embeddings and their target functions.

\section{On Centrality Measures}

In general, node-level centralities attempt to summarise a node's contribution to the network cohesion. Several centralities have been proposed and many models and  interpretations have been suggested, namely: autonomy, control, risk, exposure, influence, etc. \cite{borgatti2006graph}. Despite their myriad of applications and interpretations, in order to calculate some of these centralities one may face both  high time and space complexity, thus making it costly to compute them on large networks. Although some studies pointed out a high degree of correlation between some of the most common centralities \cite{lee2006correlations}, it is also stated that these correlations are attached to the underlying network structure and thus may vary across different network distributions \cite{schoch2017corr}. Therefore, techniques to allow faster centrality computation are topics of active research \cite{grando2018machine}.

Here, however, we are not concerned as much with the time complexity of computing the centrality measure per-se, but with whether a neural network can infer a node's centrality solely from the network structure -- that is, without any numeric information about the node, its neighbourhood, or the network itself -- even if the complexity of the methods presented here is similar to that of matrix operations, as the underlying procedures are based on these, and is polynomial with the size of the input. With this in mind, we selected four well-known node centralities to investigate in our study: \textbf{degree} -- first proposed by \cite{shaw1954degree}, it simply calculates to how many neighbours a node is connected; \textbf{betweenness} -- it calculates the number of shortest paths which cross by the given node. High betweenness nodes are more important to the graph's cohesion, i.e., their removal may disconnect the graph. A fast algorithm version was introduced by \cite{brandes2001bet}; \textbf{closeness} -- as defined by \cite{beauchamp1965clo}, it is a distance-based centrality which measures the average geodesic distance between a given node and all other reachable nodes; \textbf{eigenvector} -- this centrality uses the largest eigenvalue of the adjacency matrix to compute its eigenvector \cite{bonacich1987eig} and assigns to each node a score based upon the score of the nodes to whom it is connected. It is usually computed via a power iteration method with no convergence guaranteed.

\section{A GNN Model for Learning Relations Between Centrality Measures}

On a conceptual level, the GNN application considered here assigns multidimensional embeddings $\in \mathbb{R}^d$ to each vertex in the input graph. These embeddings are refined through $t_{max}$ iterations of message-passing. At each iteration, each vertex adds up all the messages received along its edges and adds up all the messages received along its outcoming edges, obtaining two $\mathbb{R}^d$ tensors. These two tensors are concatenated to obtain a $\mathbb{R}^{2d}$ tensor, which is fed to a Recurrent Neural Network (RNN) which updates the embedding of the vertex in question. Note that a ``message'' sent by a vertex embedding in this sense is the output of a Multilayer Perceptron (MLP) which is fed with the embedding of the vertex in question.

In summary, these models can be seen as a message-passing algorithm in which the update ($V_u$) and message-computing ($src_{msg} : \mathbb{R}^d \rightarrow \mathbb{R}^d$, $tgt_{msg} : \mathbb{R}^d \rightarrow \mathbb{R}^d$) modules are trained neural networks. With this setup we tested two different methods of extracting the centrality measures from the propagated embeddings. Our baseline method is the straightforward application of the methods commonly proposed in the GNN literature, where a MLP $approx_c : \mathbb{R}^{d} \rightarrow \mathbb{R}$, which tries to approximate each centrality measure $c$ in question directly, is trained. In the method we propose here, however, we train a MLP $cmp_c : \mathbb{R}^{2d} \rightarrow \mathbb{R}$, which is assigned with computing the probability that $v_i >_c v_j$ given their embeddings, where $>_c$ here denotes the total ordering imposed by the centrality measure $c$, that is, the node $v_i$ whose embedding is on the first $d$ dimensions has a strictly higher $c$-centrality than $v_j$ whose embedding is on the last $d$ dimensions of the input to $cmp_c$).

The first method computes the centralities directly, minimising the mean squared error (MSE) between the prediction and the true value of the centrality, and a ranking can be extracted from these directly. For such method we considered three different setups, in one we learn the normalised centrality measures directly (named AN, for ``Approximate the Normalised centrality''), in the second one we learn the unnormalised version of the centrality measures (called AU, for ``Approximate the Unnormalised centrality''), and a third approach was to learn from the normalised centrality values, but perform a normalisation of the model's approximated value before using it as its final output (dubbed AM, for ``Approximate the normalised centrality, with normalisation on the Model''). We perform Stochastic Gradient Descent (SGD), more specifically TensorFlow's Adam implementation, on the MSE loss, and evaluate these models using both absolute and relative errors, as well as computing the Kendall-$\tau$ correlation coefficient for the set of predictions in each graph.

In our proposed method, however, we only have one model (that we name as RN, for ``Rank centralities Natively by comparison'') in which for each pair of vertices $(v_i, v_j) \in \mathcal{V} \times \mathcal{V}$ and for each centrality $c \in \mathcal{C}$, our network guesses the probability that $v_i >_{c} v_j$. To train such a network we perform Stochastic Gradient Descent (SGD) on the binary cross entropy loss between the probabilities computed by the network and the binary ``labels'' obtained from the total ordering provided by $c$. This process can be made simple by organising the $n^2$ network outputs for each centrality, as well as the corresponding labels, into $n \times n$ matrices, as Figure~\ref{fig:train_matrix} exemplifies. In such a matrix, the binary cross entropy is computed as $H(\mathbf{M_{\gtrapprox_{c}}},\mathbf{T}) = -\sum_{i,j}{P(v_i >_c v_j) \log{T_{ij}}}$. Arranging the labels and cross entropy in such a way allows one to compute an accuracy on these binary predictions, which can be seen as a more strict version of the Kendall-$\tau$ correlation coefficient, since this accuracy metric penalises ties in the ranking as much as discordant rankings. With this in mind, we focus on this accuracy metric when comparing the baseline models (AN,AU,AM) with the RN model. We also provide other common metrics, calculated in the same fashion, such as Precision, Recall and True Negative rates, for the sake of completeness.

\begin{figure}[h!]
\centering
\small
        \begin{tikzpicture}[>=latex,scale=0.9, every node/.style={scale=0.9}]
            \matrix (A) [matrix of math nodes,%
                         nodes = {node style ge},%
                         left delimiter  = (,%
                         right delimiter = )] at (-0.5,0)
            {%
              P(v_1 >_c v_1) & P(v_2 >_c v_1) & P(v_3 >_c v_1) \\
              P(v_1 >_c v_2) & P(v_2 >_c v_2) & P(v_3 >_c v_2) \\
              P(v_1 >_c v_3) & P(v_2 >_c v_3) & P(v_3 >_c v_3) \\
            };
            
            \matrix (B) [matrix of math nodes,%
                         nodes = {node style ge},%
                         left delimiter  = (,%
                         right delimiter = )] at (4,0)
            {%
              0 & 1 & 1 \\
              0 & 0 & 1 \\
              0 & 0 & 0 \\
            };
        \end{tikzpicture}
    \caption{Example of a predicted fuzzy comparison matrix $\mathbf{M_{\gtrapprox_{c}}}$ at the left and the training label given by an upper triangular matrix $\mathbf{T}$ at the right, for a graph with three vertices $\mathcal{V} = \{v_1, v_2, v_3\}$ sorted in ascending centrality order as given by the centrality measure $c$. The binary cross entropy is computed as $H(\mathbf{M_{\gtrapprox_{c}}},\mathbf{T}) = -\sum_{i,j}{P(v_i >_c v_j) \log{T_{ij}}}$}.
    \label{fig:train_matrix}
\normalsize
\end{figure}

In summary, the models used here are the three baseline models, which are identical in structure, and whose only difference is the function they optimise: the AN model optimises $MSE( pred_c, norm(c) )$, the AM model optimises \\ $MSE( norm(pred_c), norm(c) )$, while the AU model optimises $MSE( pred_c, c )$. In contrast, our proposed model learns a different function and computes the rankings of the centralities through a learned native comparison on the internal embedding representation of nodes, using the cross entropy on the binary comparison between all pairs of embeddings as a loss. A complete description of our RN algorithm is presented in Algorithm~\ref{alg:GNN-centrality}.

\begin{algorithm}[h!]
\caption{Graph Neural Network Centrality Predictor}\label{alg:GNN-centrality}
\small
\begin{algorithmic}[1]
\Procedure{GNN-Centrality}{$\mathcal{G} = (\mathcal{V},\mathcal{E}), \mathcal{C}$}

\State
\LineComment{{\small Compute adj. matrix}}
\State $\mathbf{M}[i,j] \leftarrow 1 \textrm{ if } (v_i,v_j) \in \mathcal{E} \textrm{ else } 0$
\LineComment{{Initialise all vertex embeddings with the initial embedding $V_{init}$ (this initial embedding is a parameter learned by the model)}}
\State $V^{1}[i,:] \leftarrow V_{init} ~|~ \forall v_i \in \mathcal{V}$
\LineComment{Run $t_{max}$ message-passing iterations}
\For{$t=1 \dots t_{max}$}
  \LineComment{{\small Refine each vertex embedding with messages received from incoming edges}}
  \LineComment{{\small either as a source or a target vertex}}
  \label{alg:line:vertices_refinement}\State $\mathbf{V}^{t+1}, \mathbf{V}_h^{t+1} \leftarrow V_u( \mathbf{V}^t, \mathbf{M} \times src_{msg}(\mathbf{V}^t), \mathbf{M}^{T} \times tgt_{msg}(\mathbf{V}^t) )$
\EndFor

\State

\For{$c \in \mathcal{C}$}
    \LineComment{Compute a fuzzy comparison matrix $M_{\gtrapprox_{c}} \in \mathbb{R}^{|\mathcal{V}| \times |\mathcal{V}|}$}
    \State $\mathbf{M_{\gtrapprox_{c}}}[i,j] \leftarrow cmp_{c}(\mathbf{V}^{t_{max}}[i,:], \mathbf{V}^{t_{max}}[j,:]) ~|~ \forall~ v_i, v_j \in \mathcal{V}$
    \LineComment{Compute a strict comparison matrix $M_{<_{c}} \in \{\top,\bot\}^{|\mathcal{V}| \times |\mathcal{V}|}$}
    \State $\mathbf{M_{>_{c}}} \leftarrow M_{\gtrapprox_{c}} > \frac{1}{2}$
\EndFor

\EndProcedure
\end{algorithmic}
\normalsize
\end{algorithm}

\section{Experimental Setup}

We generated a training dataset by producing $4096$ graphs between $32$ and $128$ vertices for each of the four following random graph distributions (total $16384$): (1) Erdős-Rényi \cite{batagelj2005efficient}, (2) Random power law tree\footnote{This refers to a tree with a power law degree distribution specified by the parameter $\gamma$}, (3) Connected Watts-Strogatz small-world model \cite{watts1998collective}, (4) Holme-Kim model \cite{holme2002growing}. Further details are reported in Table~\ref{tab:train-inst-par}. We use 5 different datasets for evaluating our models, called ``test'', ``large'', ``different'', ``sizes'' and ``real''. The ``test'' dataset is composed of new instances sampled from the same distributions as the training dataset, and has the same number of instances. The ``large'' dataset has fewer instances, sampled from the same distributions but the number of nodes in the networks are counted between 128 and 512 vertices. The ``different'' dataset has the same number of nodes as in the training data, but these are sampled from two different distributions, namely the Barabási-Albert model \cite{albert2002statistical} with  and shell graphs\footnote{The shell graphs used here were generated with the number of points on each shell proportional to the ``radius'' of that shell. I.E., $n_i \approx \pi \times i$ with $n_i$ being the number of nodes in the i-th shell.} \cite{sethuraman2000graceful}. The ``sizes'' dataset again uses the same distributions seen during training, but with the specific sizes on a range from $32$ to $256$ with strides of $16$ to allow us to analyse the generalisation to larger instance sizes and how the performance of the models vary with the graph's size, beginning in the range with which the model was trained up until the size of the largest instances of the ``large'' dataset.  All these graphs were generated with the Python NetworkX package \footnote{\url{https://networkx.github.io/}}.
\cite{hagberg2008exploring}.

\begin{table}[h!]
    \centering
    \begin{tabular}{cc}
        \toprule
         Graph Distribution & Parameters  \\
         \midrule
         Erdős-Rényi & $p = 0.25$ \\
         Random power law tree & $\gamma = 3$ \\
         Watts-Strogatz & $k = 4, p = 0.25$ \\
         Holme-Kim & $m = 4, p = 0.1$ \\
         \bottomrule
    \end{tabular}
    \caption{Training instance generation parameters}
    \label{tab:train-inst-par}
\end{table}

The ``real'' dataset consists of real instances obtained from either the Network Repository \footnote{\url{http://networkrepository.com/}} or from the Stanford Large Network Dataset Collection \footnote{\url{https://snap.stanford.edu/data/}}, which were \emph{power-eris1176}, a power grid network, \emph{econ-mahindas}, an economic network, \emph{socfb-haverford76} and \emph{ego-Facebook}, Facebook networks, \emph{bio-SC-GT}, a biological network, and \emph{ca-GrQc}, a scientific collaboration network. The size range of these networks significantly surpass that of the trainig instances, overestimating from $\times 9$ to $\times 31$ the size of the largest ($n=128$) networks which will be seen during training, while also pertaining to entirely different graph distributions than those described in Table~\ref{tab:train-inst-par}. 

In the results presented here we instantiate our models with size $d=64$ vertex embeddings, using a simple layer-norm LSTM with $d$-dimensional output and hidden state, as well as three-layered $(d,d,d)$ MLPs $src_{msg}$, $tgt_{msg}$, $(d,d,1)$ and $(2d,2d,1)$ MLPs for the $approx_c$ and $cmp_c$ functions $\forall c \in \mathcal{C}$~\footnote{$\mathcal{C}$ here denotes the set of centrality measures}. Rectified Linear Units (ReLU) were used for all hidden layers as non-linearities and the output layer was linear, the LSTM used ReLUs instead of the traditional hyperbolic tangent. The embedding dimensionality of 64 was chosen since it improved performance over lower dimensional embeddings, no other hyperparameters were optimised.

The message-passing kernel weights are initialised with TensorFlow's Xavier initialisation method described in \cite{glorot2010understanding} and the biases are initialised to zeroes. The LSTM assigned with updating embeddings has both its kernel weights and biases initialised with TensorFlow's Glorot Uniform Initialiser \cite{glorot2010understanding}, with the addition that the forget gate bias had 1 added to it. The number of message-passing timesteps is set at $t_{max} = 32$.

Each training epoch is composed by $32$ SGD operations on batches of size $32$, randomly sampled from the training dataset (The sampling inhibited duplicates in an epoch, but duplicates are allowed across epochs). Instances were batched together by performing a disjoint union on the graphs, producing a single graph with every graph being a disjoint subgraph of the batch-graph, in this way the messages from one graph were not be passed to another, effectively separating them.

All models were tested both by training a single model for each centrality as well as building a model which shared the computed node embedding over different centralities, learning only a different $approx_c$ or $cmp_c$ function as necessary for each centrality. This allows us to measure whether such multitask learning of different centrality measures has any impact, positive or negative, on the model's performance.

\section{Experimental Analysis}

After 32 training epochs, the models were able to compute centrality predictions and comparisons. The approximation models had high errors for some of the centralities, but the Kendall-$\tau$ correlation coefficient on the AN and AM models was satisfactory when calculated on the generated datasets, results similar to those reported in \cite{grando2015estimating,grando2016approximating,grando2018computing,grando2018machine}. The AU model had a low Kendall-$\tau$ correlation coefficient for the multitasking model, even for the ``test'' dataset, which mirrors how the normalisation for each centrality allows the network to perform better due to a common range of possible values.

Thus, we turned ourselves to comparing the three remaining models which presented a reasonable accuracy, often achieving over $80\%$ training accuracy (averaged over all centralities). However the AM model had difficulties performing as good consistently on the ``test'' dataset, where the AN and RN models did not have any problem whatsoever. The accuracy of the multitasking AN model was consistently and significantly worse than the non-multitasking models, while in the RN model the multitasking is outperformed by the basic model in many cases, but the overall accuracy is not significantly different (see Table~\ref{tab:test-results} for further results and comparisons). In the big picture, however, the RN model consistently outperformed the others, lagging only on the Recall metric.

\begin{table*}[bt]
    \centering
    \setlength{\tabcolsep}{2.5pt}
    \begin{tabular}{|c|cccc|cccc|cccc|}
    \hline
        & & \multicolumn{2}{c}{AN} & & & \multicolumn{2}{c}{AM} & & & \multicolumn{2}{c}{RN} & \\
        Centrality & P & R & TN & Acc & P & R & TN & Acc & P & R & TN & Acc \\
        \hline
        \multicolumn{13}{c}{Without multitasking}\\
        \hline
        Betweenness & $82.4$ & $\mathbf{88.6}$ & $84.6$ & $86.0$
                        & $39.9$ & $45.6$ & $43.5$ & $44.1$
                        & $\mathbf{90.3}$ & $88.5$ & $\mathbf{91.0}$ & $\mathbf{89.8}$ \\
        Closeness   & $83.1$ & $\mathbf{85.4}$ & $84.3$ & $84.8$
                        & $83.1$ & $\mathbf{85.4}$ & $84.4$ & $84.9$
                        & $\mathbf{88.4}$ & $84.5$ & $\mathbf{89.8}$ & $\mathbf{87.3}$ \\
        Degree      & $75.3$ & $\mathbf{98.4}$ & $81.6$ & $87.5$
                        & $63.3$ & $85.8$ & $70.4$ & $75.6$
                        & $\mathbf{99.3}$ & $94.9$ & $\mathbf{99.4}$ & $\mathbf{97.6}$ \\
        Eigenvector & $87.0$ & $87.0$ & $87.6$ & $87.3$
                        & $86.5$ & $86.5$ & $87.0$ & $86.8$
                        & $86.2$ & $\mathbf{90.2}$ & $82.3$ & $86.3$ \\
        Average     & $82.0$ & $\mathbf{89.8}$ & $84.5$ & $86.4$
                        & $68.2$ & $75.8$ & $71.3$ & $72.8$
                        & $\mathbf{91.0}$ & $89.5$ & $90.6$ & $\mathbf{90.3}$ \\    
        \hline
        \multicolumn{13}{c}{With multitasking}\\
        \hline
        Betweenness & $73.6$ & $79.6$ & $76.1$ & $77.3$
                        & $41.9$ & $46.7$ & $46.3$ & $46.2$
                        & $87.2$ & $87.2$ & $88.9$ & $87.9$ \\
        Closeness   & $71.4$ & $73.0$ & $73.6$ & $73.3$
                        & $80.2$ & $82.5$ & $81.7$ & $82.1$
                        & $86.9$ & $82.0$ & $88.7$ & $85.5$ \\
        Degree      & $73.7$ & $96.0$ & $80.4$ & $85.9$
                        & $72.4$ & $94.3$ & $79.3$ & $84.7$
                        & $98.3$ & $92.4$ & $99.0$ & $96.4$ \\
        Eigenvector & $83.5$ & $83.5$ & $84.2$ & $83.9$
                        & $85.5$ & $85.5$ & $86.2$ & $85.9$
                        & $\mathbf{89.8}$ & $88.3$ & $\mathbf{90.4}$ & $\mathbf{89.4}$ \\
        Average     & $75.5$ & $83.0$ & $78.6$ & $80.1$
                        & $70.0$ & $77.3$ & $73.4$ & $74.7$
                        & $90.5$ & $87.5$ & $\mathbf{91.8}$ & $89.8$ \\
    \hline
    \end{tabular}
    \caption{Performance metrics (Precision, Recall, True Negative rate, Accuracy) computed for the trained models on the ``test'' dataset. All values are presented as percentages and the best value for each metric and centrality combination is highlighted.}
    \label{tab:test-results}
\end{table*}

In this context, recall that the multitask learning model is required to develop a ``lingua franca'' of vertex embeddings from which information about any of the studied centralities can be easily extracted, so in a sense it is solving a harder problem. We also computed performance metrics for the ``large'' dataset where the overall accuarcy of the AM model falls to about $65\%$, while it surprisingly was able to maintain performance for the eigenvector centrality. The accuracy of the AN model falls a few percent points to $81\%$ and $78\%$ (without/with multitasking), while the RN model also had its performance lowered to about $80\%$, a sharper drop but still performing on par with the AN model. This result shows that our proposed model is able to generalise to larger problem sizes than those it was trained on, with a expected decrease accuracy, while also being able of handling multitasking without interference on the performance of other centralities.

\subsection{Generalising to Other Distributions}

Having obtained good performance for the distributions the network was trained on, we wanted to assess the possibility of accurately predicting centrality comparisons for graph distributions it has never seen. That was done by computing performance metrics on the ``different'' dataset for which the results are reported in Table~\ref{tab:test-other-distr-results}. Although its accuracy is reduced in comparison ($81.9\%$ vs $90.3\%$ overall), the model can still predict centrality comparisons with high performance, obtaining its worst result at $82.0\%$ recall for the closeness centrality. The model without multitasking outperforms the multitasking one only by a narrow margin ($0.5\%$ at the overall accuracy), while the multitasking model has a better accuracy than even the best baseline model. The comparison model was only outperformed in the recall metric and only by $0.3\%$ in average.

\begin{table*}[bt]
    \centering
    \setlength{\tabcolsep}{2.5pt}
    \begin{tabular}{|c|cccc|cccc|cccc|}
    \hline
        & & \multicolumn{2}{c}{AN} & & & \multicolumn{2}{c}{AM} & & & \multicolumn{2}{c}{RN} & \\
        Centrality & P & R & TN & Acc & P & R & TN & Acc & P & R & TN & Acc \\
        \hline
        \multicolumn{13}{c}{Without multitasking}\\
        \hline
        Betweenness & $80.2$ & $\mathbf{80.3}$ & $80.8$ & $\mathbf{80.5}$
                        & $43.1$ & $43.2$ & $44.8$ & $44.0$
                        & $\mathbf{81.2}$ & $77.5$ & $\mathbf{81.8}$ & $79.7$ \\
        Closeness   & $80.8$ & $82.4$ & $81.6$ & $82.0$
                        & $80.8$ & $82.4$ & $81.6$ & $82.0$
                        & $81.7$ & $75.3$ & $\mathbf{84.2}$ & $79.9$ \\
        Degree      & $82.7$ & $94.6$ & $85.2$ & $89.0$
                        & $75.3$ & $86.8$ & $78.3$ & $81.7$
                        & $86.4$ & $72.5$ & $89.0$ & $82.1$ \\
        Eigenvector & $70.4$ & $70.4$ & $71.2$ & $70.8$
                        & $69.9$ & $69.9$ & $70.7$ & $70.3$
                        & $\mathbf{84.9}$ & $\mathbf{87.9}$ & $\mathbf{83.8}$ & $\mathbf{85.8}$ \\
        Average     & $78.5$ & $\mathbf{81.9}$ & $79.7$ & $80.6$
                        & $67.3$ & $70.6$ & $68.9$ & $69.5$
                        & $\mathbf{83.6}$ & $78.3$ & $\mathbf{84.7}$ & $\mathbf{81.9}$ \\
        \hline
        \multicolumn{13}{c}{With multitasking}\\
        \hline
        Betweenness & $73.4$ & $73.6$ & $74.3$ & $73.9$
                        & $46.6$ & $46.8$ & $48.2$ & $47.5$
                        & $77.9$ & $77.0$ & $78.5$ & $77.8$ \\
        Closeness   & $81.3$ & $82.9$ & $82.2$ & $82.5$
                        & $\mathbf{81.9}$ & $\mathbf{83.5}$ & $82.7$ & $\mathbf{83.1}$
                        & $79.6$ & $77.5$ & $81.4$ & $79.5$ \\
        Degree      & $85.0$ & $\mathbf{97.0}$ & $87.4$ & $\mathbf{91.3}$
                        & $84.1$ & $96.1$ & $86.5$ & $90.3$
                        & $\mathbf{87.4}$ & $74.9$ & $\mathbf{91.0}$ & $84.0$ \\
        Eigenvector & $67.5$ & $67.5$ & $68.4$ & $68.0$
                        & $70.4$ & $70.4$ & $71.1$ & $70.7$
                        & $79.6$ & $80.5$ & $79.9$ & $80.2$ \\
        Average     & $76.8$ & $80.3$ & $78.1$ & $78.9$
                        & $70.8$ & $74.2$ & $72.1$ & $72.9$
                        & $81.1$ & $77.5$ & $82.7$ & $80.4$ \\
    \hline
    \end{tabular}
    \caption{Performance metrics (Precision, Recall, True Negative rate, Accuracy) computed for the trained models on the ``different'' dataset. All values are presented as percentages and the best value for each metric and centrality combination is highlighted.}
    \label{tab:test-other-distr-results}
\end{table*}

\begin{figure}[h]
    \centering
    \includegraphics[width=.9\linewidth]{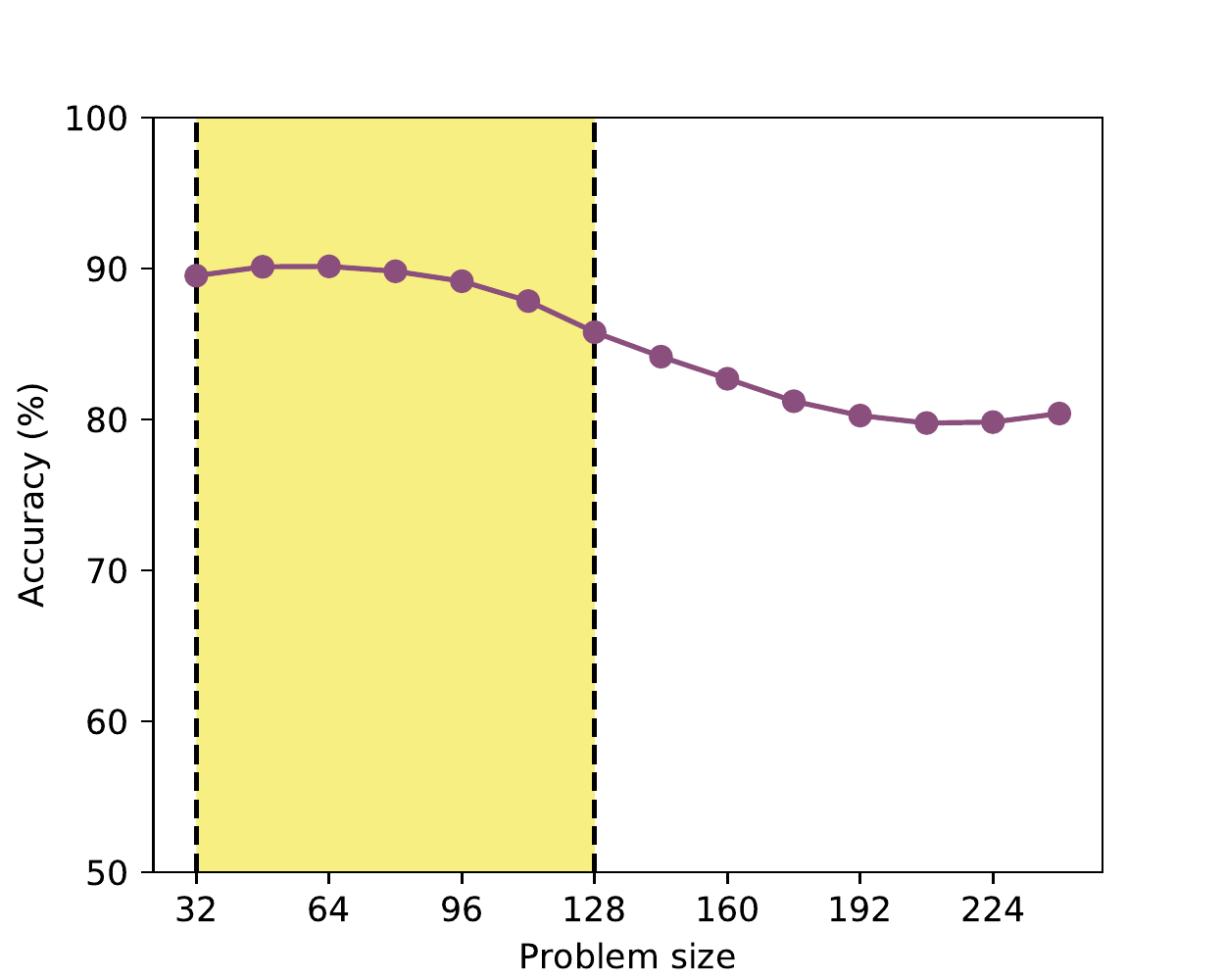}
    \caption{The overall accuracy decays with increasing problem sizes, although it still does not approach $50\%$ (equivalent to randomly guessing each vertex-to-vertex centrality comparison) for the largest instances tested here. The dotted lines delimit the range of problem sizes used to train the network ($n=32 \dots 128$).}
    \label{fig:test-var-sizes}
\end{figure}

Testing the models on the ``sizes'' dataset provides insight on how their performance decays on larger instances. Our multitask model accuracy presents an expected decay with increasing problem sizes (see Figure~\ref{fig:test-var-sizes}), while the non-multitasking models performed similarly. However, this decay is not a free fall towards $50\%$ for instances with almost twice the size of the ones used to train the model, in fact the overall accuracy remains around $80\%$ when $n=240$ which implies that some level of generalisation to even larger problem sizes is achievable.

The approximation models, however, had different behaviours for each centrality. The RM model was able to maintain its performance relatively stable for the eigenvector centrality, while with degree and closeness it presented sharp drops inside and outside the range of sizes it was trained upon (the betweenness centrality did not reach $50\%$ accuracy and was not taken into consideration). The RN model was the most stable overall, only having a significant drop in performance with the closeness centrality when tested on instances with $176$ or more nodes.

We also wanted to assess the model's performance on real world instances. We analysed the performace of the model on the ``real'' dataset. The RN model was able to obtain up to $86.04\%$ accuracy (on degree) and $81.79\%$ average accuracy on the best case (\emph{bio-SC-GT}), and $57.85\%$ accuracy (closeness) and $65.47\%$ average accuracy on the worst case (\emph{ego-Facebook}). Overall the accuracy for the models without/with multitask were $74.6\%$ and $73.8\%$, which are only slightly worse than the non-multitasking AN model, which has a $76.8\%$, owing much of this to its high performance on the degree centrality, which it predicts with almost $90\%$ accuracy over all graphs. Due to the extreme overestimation of the number of nodes and the different distributions present in these graphs, we found it impressive that these two models can predict with such high accuracies. And although our model was outclassed in the non-multitasking environment, the multitasking RN model outperforms the multitasking AN model (with $68.34\%$ accuracy) by a large margin, while being faster by computing all four centralities at once.

\subsection{Interpretability}

Machine learning has achieved impressive feats in the recent years, but the interpretability of the computation that takes place within trained models is still limited \cite{lipton2016mythos}. \cite{selsam2018learning} have shown that it is possible to extract useful information from the embeddings of CNF literals, which they manipulate to obtain satisfying assignments from the model (trained only as a classifier). This allowed them to deduce that Neurosat works by guessing UNSAT as a default and changing its prediction to SAT only upon finding a satisfying assignment.

\begin{figure*}[tb]
    \centering
    \includegraphics[width=0.9\linewidth]{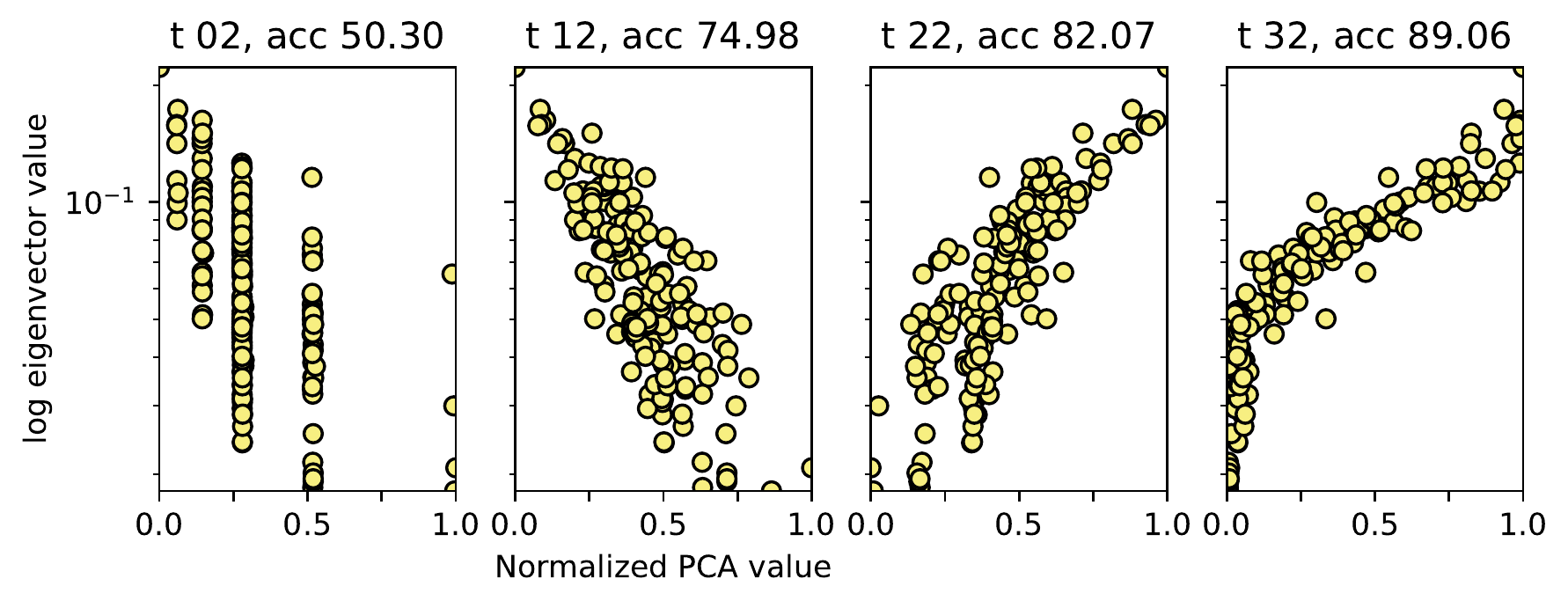}
    \caption[Plot of the evolution of the 1D PCA for the model]{Evolution of the 1D projection of vertex embeddings of a non-multitask model plotted against the corresponding eigenvector centralities (plotted on a log scale) through the message-passing iterations for a graph sampled from the Watts-Strogatz small world distribution of the ``large'' dataset.}
    \label{fig:1d-pca}
\end{figure*}

In our case, we can obtain insights about the learned algorithm by projecting the refined set of embeddings $V^{t_{max}} \in \mathbb{R}^{64}$ onto one-dimensional space by the means of Principal Component Analysis (PCA) and plotting the projections against the centrality values of the corresponding vertices. We do this only for the RN model, since the comparison framework might suggest that the learned embeddings need not represent the centrality values directly. Figure~\ref{fig:1d-pca} shows the evolution of this plot through message-passing iterations, from which we can infer some aspects of the learned algorithm. First of all, the zeroth step is suppressed due to space limitations, but because all embeddings start out the same way, it corresponds to a single vertical line. One can see how the PCA of the embeddings seem to sort the vertices along their eigenvector centrality while the model improves its performance. The cause for the most central vertex shifting from being on the left to the right is most likely due to the normalisation of the PCA projection. As the solution process progresses, the network progressively manipulates each individual embedding to produce a correlation between the centrality values and the $\mathbb{R}^d$ vector, which can be visualised here as reordering data points along the horizontal axis. 

The case shown here, however, is not universally true, and vary somewhat depending on the distribution from which the graph was drawn. Graphs sampled from the power law tree distribution, for example, seem to be more exponential in nature when comparing the log-centrality value and the normalised 1-dimensional PCA value. But most of the distributions trained on had a similar behaviour of making a line between the logarithm of the centrality and the normalised PCA values. However, even in the cases where the centrality model did not achieve a high accuracy, we can still look at the PCA values and see whether they yield a somewhat sensible answer to the problem. Thus, the embeddings generated by the network can be seen as the GNN trying to create a centrality measure of its own with parts of, or the whole embedding being correlated with those centralities with which the network was trained.

\section{Conclusions}

In this paper, we demonstrated how to train a neural network to predict graph centrality measures, while feeding it with only the raw network structure. 
In order to do so, we enforced permutation invariance among graph elements by engineering a message-passing algorithm composed of neural modules with shared weights. These modules can be assembled in different configurations to reflect the network structure of each problem instances. 
We show that our proposed model which is trained to predict centrality comparisons (i.e. is vertex $v_1$ more central than vertex $v_2$ given the centrality measure $c$?) performs slightly better than the baseline model of predicting the centrality measure directly. We draw parallels between the accuracy defined in this setup and the Kendall-$\tau$ correlation coefficient, and show that our model performs reasonably well within this metric and that this performance generalises reasonably well to other problem distributions and larger problem sizes. We also show that the model shows promising performance for very large real world instances, which overestimate the largest instances known at training time from $\times$9 to $\times$31 (4,000 as opposed to 128 vertices). We also show that although our model can be instantiated separately for each centrality measure, it can also be trained to predict all centralities simultaneously, with minimal effect to the overall accuracy.

The model presented here however, isn't without its failings. Some models in the literature are able to provide meaningful node embeddings with lower dimensionality values than the ones we present here, however their work does not focus on centrality measures and some use information about the graph other than its topology, as well as learn the embeddings for similar graph distributions or even the same graph used for training. Although our model is presented here with four sample centralities, some of which are easier to compute precisely rather than approximating with our method, we believe that the joint prediction of multiple centralities is still useful, and that the time complexity of our model, which is dependant mostly on matrix multiplications, still remains polynomial even if one increases the amount of centralities being predicted.

In summary, this work presents, to the best of our knowledge, the first application of Graph Neural Networks to multiple centrality measures and the proposal of a comparison framework for processing node embeddings. We yield an effective model and provide ways to have such a model work with various centralities at once, in a more memory-efficient way than having a different model for every centrality -- with minimal loss in its performance. Finally, our work attests to the power of relational inductive bias in neural networks, allowing them to tackle graph-based problems, and also showing how the proposed model can be used to provide a network that condenses multiple information about a graph in a single embedding.

\section*{Acknowledgements}

This study was financed in part by the Coordenação de Aperfeiçoamento de Pessoal de Nível Superior - Brasil (CAPES) - Finance Code 001, and by the Brazilian Research Council CNPq. We gratefully acknowledge the support of NVIDIA Corporation with the donation of the Quadro P6000 GPU used for this research.

\bibliographystyle{plain}
\bibliography{gnncentrality}

\end{document}